\documentclass[a4paper, 12pt]{article}
\usepackage{amsmath}
\usepackage{amssymb}
\usepackage[utf8]{inputenc}
\usepackage{graphicx}
\usepackage{subfigure}
\usepackage[font=small, skip=0pt]{caption}
\usepackage{cleveref}
\def \bea {\begin{eqnarray}}  
\def \eea {\end{eqnarray}} 
\def \mea {\nonumber\\}
\def \half {{\textstyle \frac{1}{2}}}

\begin{document}    
\begin{titlepage}

\title{Solutions of Painlev\'e II on real intervals: novel approximating sequences}     
     
\author{ 
A.J. Bracken\footnote{{\em Email:} a.bracken@uq.edu.au}
\\School of Mathematics and Physics\\    
The University of Queensland\\Brisbane 4072, Australia}

\date{}     
\maketitle 
\begin{abstract} 
Novel sequences of approximants to solutions of Painlev\'e II on finite intervals of the real line, 
with Neumann boundary conditions, are constructed.   Numerical experiments strongly suggest convergence of these sequences 
in a surprisingly wide range of cases, even ones where ordinary perturbation series fail to converge.  
These sequences are here labeled extraordinary because of their unusual properties.
Each  
element of such a  
sequence is defined on its own interval.  As the sequence  (apparently) converges
to a solution of the corresponding boundary value problem for Painlev\'e II, these intervals 
themselves (apparently) converge to the defining interval for that problem, 
and an associated sequence of constants (apparently) converges
to the constant term in the Painlev\'e II equation itself.    Each extraordinary sequence is
constructed in a nonlinear fashion from a perturbation series approximation to the solution of   a
supplementary boundary value problem, involving a generalization of Painlev\'e II 
that arises in studies of electrodiffusion.   
\end{abstract}

\end{titlepage}
\setcounter{page}{2}

\section{Introduction} 
Painlev\'e's second nonlinear ordinary differential equation PII [Painlev\'e 02] continues to be 
studied widely, not only because of its intrinsic mathematical significance (see 
[Ablowitz and Segur 77; Joshi and Kruskal 94;
Bassom et al. 98;  Sakai 01; Clarkson 06; Kajiwara et al. 17] and references therein),
but also because of
its association with  nonlinear systems of interest in applications
(see
[Bass 64; Rogers et al. 99; Zaltzman and Rubinstein 07; Bass et al. 10; Bracken and Bass 18] and references therein).
Two families of  solutions of PII are known in special cases  [Bass et al. 10], expressed in terms of ratios of polynomials 
in the one [Yablonskii 59; Vorob'ev 65],  and ratios of 
Airy functions in the other [Lukashevich 71; Clarkson 16],
but in general solutions have not been expressed in terms of more familiar functions and are
referred to as  
Painlev\'e transcendents [Ince 56; Clarkson 06].

Consider the boundary value problem (BVP) for PII in standard form on the real  interval $[a,\,b]$, with
Neumann boundary conditions (BCs), defined by
\bea
y''(z)=2y(z)^3+zy(z)+C \,,\quad a<z<b\,,\quad  y'(a)=0=y'(b)\,.
\label{painleve1}
\eea
In particular, consider the case where
\bea
a= -15.650\dots\,,\quad b= -14.911\dots\,,\quad C= -1.468\dots
\label{abc1}
\eea
The reasons for this choice of values  will be made clear below, 
where it will also be explained  why they are only numerically determined, and how we  know
that in this case there exists a solution $y(z)$ that is
free from singularities, monotonically decreasing and everywhere positive
on the interval of interest. 

In what follows it is shown that for this particular BVP and uncountably many  others
of the form \eqref{painleve1}, there exists a novel type of approximating sequence, to be denoted
by
$y_E^{(n)}(z)$,  $n=1\,,2\,,\dots$,  which  appears to converge to a solution $y(z)$.
For the example defined by \eqref{painleve1} and \eqref{abc1}, 
Fig. 1 on the left and right shows  plots of $y_E^{(n)}(z)$ for  $n=1\,,2\,,3,\,4$ and for 
$n=8,\,9,\,10,\,11$, respectively, together with $y(z)$.

\begin{figure}[h]
\centering
\includegraphics[trim=0.1in 3.0in 0.1in 3.5in, clip, scale=0.6]{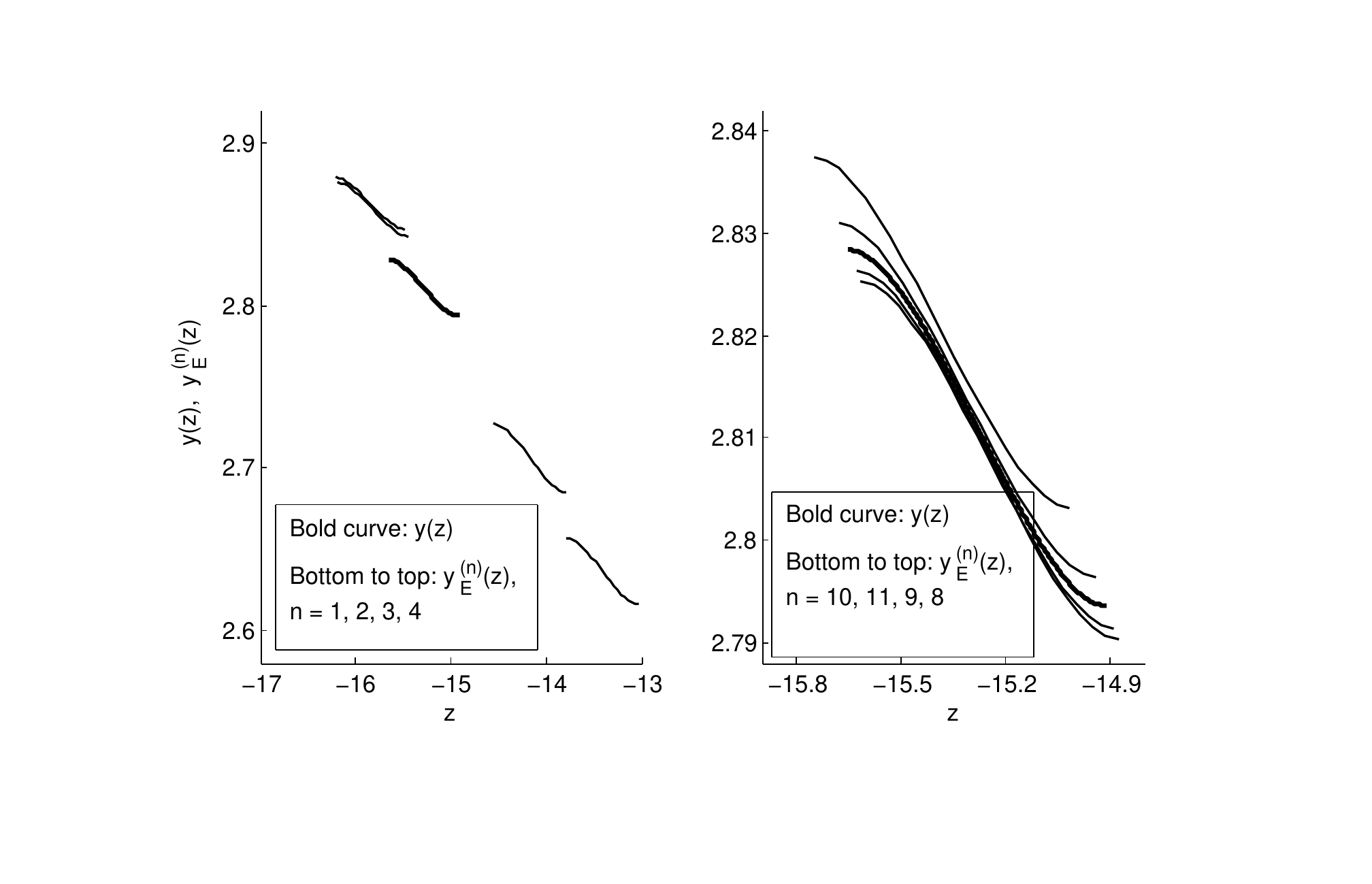}
\caption{Plots of $y(z)$ and selected $y_E^{(n)}(z)$, suggesting convergence.}
\end{figure}

This  type of sequence is highly unusual, and henceforth it will be labeled  {\em extraordinary}.
In the first place, in such a sequence each approximant $y_E^{(n)}(z)$ 
is defined on its own interval $[a_n\,,b_n]$, and is accompanied by a constant $C_n$.  In the (apparently) convergent
cases, the $a_n$, $b_n$ and $C_n$ values converge to values $a$, $b$ and $C$ while the approximants converge to
a solution of \eqref{painleve1} with these limiting parameter values. This  behaviour 
is partially evident in Fig. 1.  
To make it clearer, Fig. 2 on the left
shows the progression of the values $a_n$ and $b_n$  towards $a$ and $b$, while Fig. 2 on the right 
shows the progression of
$C_n$ values towards $C$, with $a$, $b$, and $C$ as in \eqref{abc1}.  

\begin{figure}[h]
\centering
\includegraphics[trim=0.1in 3.0in 0.1in 3.5in, clip, scale=0.6]{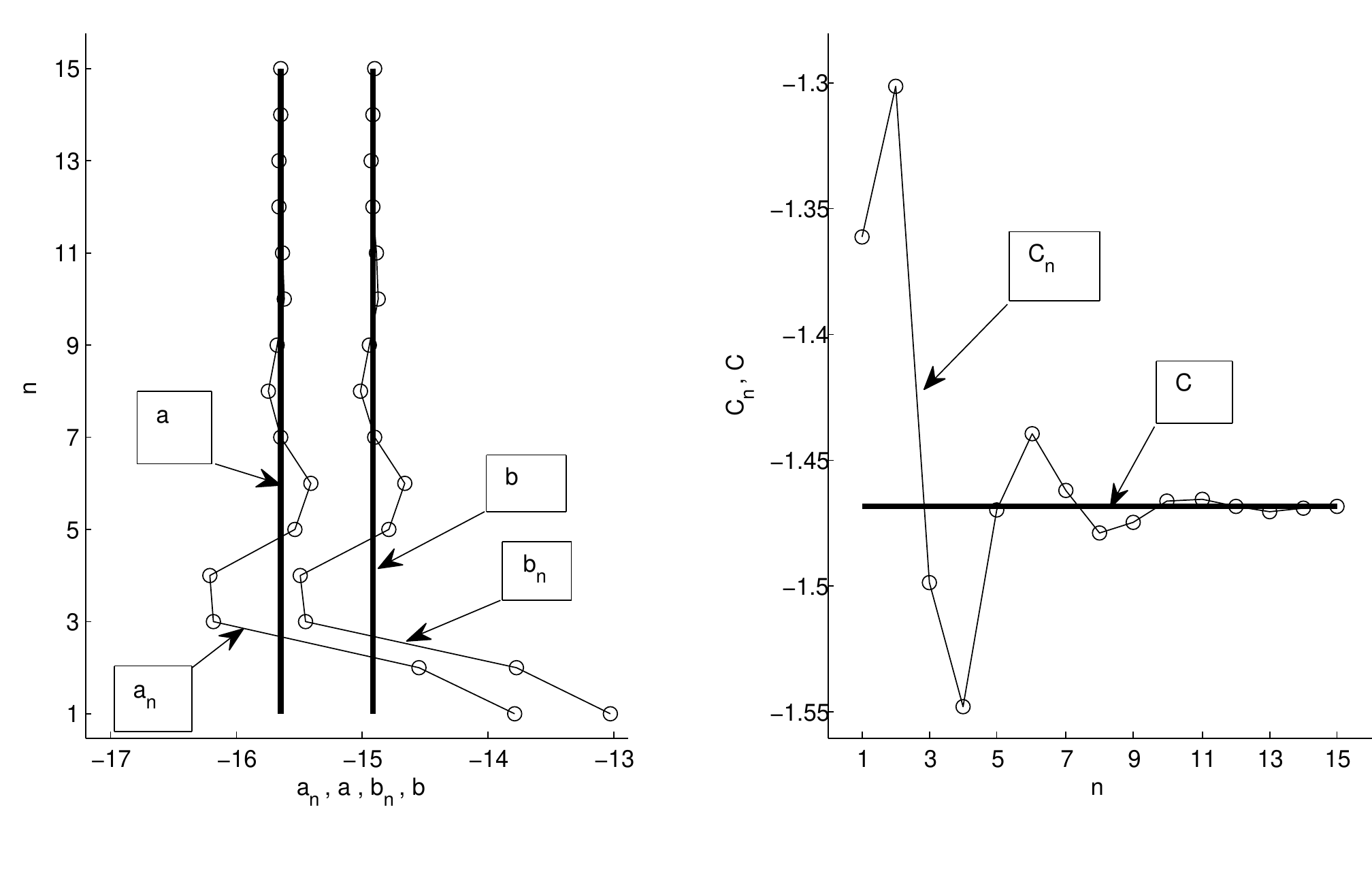}
\caption{Plots of $a_n$, $b_n$ and $C_n$ versus $n$, showing convergence to $a$, $b$ and $C$ as in \eqref{abc1}.}
\end{figure}

In the second place, the values $a_n$, $b_n$ and $C_n$ at each stage, as well as their limiting values $a$, $b$ and $C$,
can only be estimated numerically, and are not known exactly {\em a priori}.  

In the third place, and perhaps most remarkable of all, is the complicated way in which such an 
extraordinary sequence is constructed.  
Perturbation theory is first applied to a different, supplementary  BVP --- one that
 involves a generalization 
of PII arising in an application to electrodiffusion --- to obtain an (ordinary) sequence of approximants 
to the solution of that supplementary
BVP.  A nonlinear procedure is then applied to convert this ordinary sequence  into an extraordinary sequence of 
approximants to a solution of a BVP \eqref{painleve1} involving PII.   
It does not seem to be possible to construct
such extraordinary sequences directly from BVPs of the form \eqref{painleve1}.

\section{Constructing  extraordinary sequences}
The  supplementary BVP to be used,
derives from a system of coupled first-order nonlinear ODEs
 that govern a model of two-ion electrodiffusion.  The model and the ensuing BVP have
been known for over $50$ years [Bass 64],
but have been the subject of renewed interest
and much associated research in
recent years 
[Rogers et al. 99;
Bass et al. 10; Amster et al. 11; 
Bracken at al. 12; Bass and Bracken 14;  Bracken and Bass 16;  Bracken and Bass 18].

 The BVP consists of the 
 ODE
\bea 
2\nu E\,''(x)
=\nu E(x)^3 +\left\{4\sigma+\nu [E(0)^2-E(1)^2]\right\}xE(x)
\mea\mea
+\left\{2-2\sigma
-\nu E(0)^2\right\}E(x)
+\nu \tau \left\{E(0)^2-E(1)^2\right\}-4\mu
\label{painleve2}
\eea
for $0<x<1$, with Neumann BCs
\bea
E\,'(0)=0=E\,'(1)\,.
\label{BCs2}
\eea
The parameters
$\nu$, $\sigma$, $\tau $ and $ \mu$ appearing here take any chosen constant values, subject to
\bea
0<\nu\,,\quad 0<\sigma<1\,,\quad 
 -1<\tau<1\,,\quad -\infty<\mu<\infty\,.
\label{parameters1}
\eea
Their interpretation  in the context of electrodiffusion and that of the variables $x$ and $E(x)$  
appearing in \eqref{painleve2},  is not important
to the present work.  Interested
readers are refererred to earlier works,
in particular to Eqn. (14) of [Bracken and Bass 18],
where $\sigma$, $\tau$ and $\mu$ are denoted
 $1-2c_0 (=2c_1-1)$, $1-2\tau_+ (=2\tau_--1)$ and $j_1$, respectively, because of that interpretation.

Note that as well as involving a cubic nonlinearity and an $x$-dependent term similar to those found in PII in standard form \eqref{painleve1}, the ODE \eqref{painleve2} has the unusual complicating feature that 
the unknown values $E(0)$ and $E(1)$ appear nonlinearly in it, and must be
found together with  $E(x)$ for $0<x<1$, as part of any solution.  Despite this complication, 
it is known [Thompson 94; Park and Jerome  97;  Amster et al. 11; Bracken et al. 12; 
Bracken and Bass 16; Bracken and Bass 18]
that there exists a unique solution
of the BVP involving this generalized version  \eqref{painleve2} of PII,  and
that this solution is singularity-free and either 
monotonically decreasing and positive (Type A), or monotonically increasing and negative (Type B),  
everywhere on the interval $[0,\,1]$.  From this
it follows in particular that in every solution 
\bea
E(0)^2>E(1)^2\,.
\label{BCs3}
\eea

A series expansion of $E(x)$ satisfying \eqref{painleve2} and \eqref{BCs2}, including 
the end point values $E(0)$ and $E(1)$, has been obtained [Bracken and Bass 18] by perturbing away from the 
solution $E(x)=0$ when $\mu=0$ in \eqref{painleve2}.  Numerical experiments in that work strongly suggest
convergence of the series for a wide range values of the parameters \eqref{parameters1},
much wider than expected on the basis of earlier studies [Bass 64] and approximation schemes [MacGillivray 68], 
and
in particular for cases with $\sigma=1/3$, $\tau=-0.2$,
and 
\bea
0< \nu\leq 10\,,\quad -2<\mu<2\,.
\label{parameters2}
\eea   
(For values of $|\mu|$ greater than about $2$, the series typically 
appears to diverge, whatever the values of the other parameters.) 

To summarize the method, the series expansion is obtained 
by introducing a book-keeping parameter $\epsilon$  that can later be set equal to $1$, 
replacing
$\mu$ by $\epsilon \mu$  in \eqref{painleve2}, and seeking the solution in the form
\bea
E(x)=0+\epsilon E_1(x)+\epsilon^2 E_2(x)+\dots\,,
\label{Eseries1}
\eea
while emphasizing that such an expansion is also applied to the end-point values  $E(0)$ and $E(1)$ 
appearing in \eqref{painleve2}.
 Equating terms of the same degree in $\epsilon$,  it is found that $E_n(x)$ satisfies
 \bea
 \nu E_n\,''(x)= (1-\sigma+2\sigma x)\,E_n(x)+R_n(x)\,,\quad  E_n\,'(0)=0=E_n\,'(1)\,,
 \label{airy1}
 \eea
 where $R_n(x)$ depends only on the $E_k(x)$, $k=1\,,2\,,\dots \,,n-1$, including their end-point values. (For further details, 
 see Sec. 4 in [Bracken and Bass 18].)

Linearly-independent solutions of the homogeneous ODE $(R_n=0)$ in \eqref{airy1} are provided by 
\bea
A(x)={\rm Ai}(s)\,,\quad B(x)={\rm Bi}(s)\,,\quad s=(1-\sigma +2\sigma x)/(4\nu\sigma^2)^{1/3}\,,
\label{airy2}
\eea
where Ai and Bi are Airy functions of the first and second kind
[Abramowitz and Stegun 64].
Because  the
Wronskian of Ai and Bi is given by $1/\pi$, that of  $A$ and $B$ is given by 
\bea
W= [2\sigma/(\pi^3\nu)]^{1/3}\,,
\label{airy3}
\eea
and  the method of variation of parameters gives the general solution of the ODE in \eqref{airy1} as
\bea
E_n(x)=-\frac{1}{\nu W}\,\left\{A(x)\int_0^{x}R_n(y)B(y)\,dy - B(x)\int_0^{x}R_n(y)A(y)\,dy\right\}
\mea\mea
+ d_{n,A}\,A(x)+d_{n,B}\,B(x)\qquad\qquad
\label{airy4}
\eea
with $d_{n,A}$, $d_{n,B}$ arbitrary constants.  
Imposing the BCs in  \eqref{airy1} then gives 
\bea
d_{n,A}=\frac{B'(0)}{[A'(1)B'(0)-A'(0)B'(1)]\,\nu W}\left\{A'(1)\int_0^1 R_n(y)B(y)\,dy \right.
\mea\mea
\left. - B'(1)\int_0^1 R_n(y)A(y)\,dy\right\}\,,
\mea\mea
d_{n,B}=-A'(0)\,d_{n,A}/B'(0)\,.\qquad\qquad\qquad
\label{airy5}
\eea
 It may be noted in passing that the first non-zero term $E_1(x)$ in \eqref{Eseries1} was found 
 many years ago
 [Bass 64] 
 in the form \eqref{airy4}
as the solution of the linearised version of \eqref{painleve2}.  
 
The next step is to set 
\bea
E^{(n)}(x)=\sum_{k=1}^{n}\, E_k(x)\,,\quad n=1\,,2\,,\dots
\label{Esequence1}
\eea
so defining a sequence of approximants to $E(x)$.  In particular, this defines  $n$th approximations
$E^{(n)}(0)$ to $E(0)$ and $E^{(n)}(1)$ to $E(1)$.  The accuracy of the  $n$th
approximation  is tested by introducing the error measure
\bea
\Delta_n={\rm max}\,\left\{\left|E^{(n)}(x)- E(x)\right| 
+ 
\left|E^{(n)}\,'(x)-E\,'(x)\right|\right\} 
\label{Delta_def}
\eea
over all $x\in [0\,,1]$.

Critically important in what follows is that $E(x)$ satisfying \eqref{painleve2} 
can be converted by a nonlinear transformation involving $E(0)$ and $E(1)$
into a solution $y(z)$ of a corresponding BVP  \eqref{painleve1} for PII 
[Bass 64].
The conversion is defined by formulas (3.9) -- (3.12) 
in [Bass et al. 10] and formulas (13) in [Bracken and Bass 18], which set
\bea
y(z)=\frac{1}{2\beta}\,E\left(\frac{z-\gamma}{\beta}\right)\,,\quad \gamma=a\leq z\leq b=\gamma +\beta\,,
\mea\mea
C=\frac{\nu\tau\left[E(0)^2-E(1)^2\right]-4\mu}{4\nu\beta^3}\,,\qquad
\label{conversion1}
\eea
where
\bea
\beta=\left(\frac{2\sigma}{\nu}+\half\left[E(0)^2-E(1)^2\right]\right)^{1/3}\,,
\quad
\gamma=\frac{1}{\nu\beta^2}\,\left[1-\sigma-\half\nu E(0)^2\right]\,.
\label{conversion2}
\eea
Note that $\beta>0$ as a consequence of \eqref{parameters1} and \eqref{BCs3} (implying $b>a$), and also 
from \eqref{conversion1} that
the monotonicity and definite sign of the solution $E(x)$ translates into similar properties for $y(z)$.   

Corresponding to \eqref{Esequence1}, a sequence of approximants $y_E^{(n)}(z)$ to $y(z)$ is obtained 
in the form
\bea
y_E^{(n)} (z)=\frac{1}{2\beta_n}\,E^{(n)}\left(\frac{z-\gamma_n}{\beta_n}\right)\,,
\quad \gamma_n=a_n\leq z\leq b_n=\gamma_n+\beta_n\,,
\label{conversion3}
\eea
with
\bea
\beta_n=\left(\frac{2\sigma}{\nu}+\half\left[E^{(n)}(0)^2-E^{(n)}(1)^2\right]\right)^{1/3}\,,
\mea\mea
\gamma_n=\frac{1}{\nu\beta_n\,^2}\,\left[1-\sigma-\half\nu E^{(n)}(0)^2\right]\,.\qquad
\label{conversion4}
\eea
As a consequence of the BCs  \eqref{BCs2}, these approximants satisfy
\bea
y_E^{(n)}\,'(a_n)=0=y_E^{(n)}\,'(b_n)\,.
\label{conversion5a}
\eea
Thus the function $y(z)$  is (potentially) approached by a sequence of approximants
$y_E^{(n)}(z)$ defined on intervals $[a_n,\,b_n]$ that can differ from one value of $n$ to the next  
and that approach an interval $[a,\,b]$ determined only 
as $n\to \infty$.  
Furthermore, corresponding to the definition of $C$ in \eqref{conversion1}, we have
\bea
C_n=\frac{\nu\tau\left[E^{(n)}(0)^2-E^{(n)}(1)^2\right]-4\mu}{4\nu\beta_n \,\!^3}\,,
\label{conversion5}
\eea 
so that the value of $C$ in the ODE \eqref{painleve1} satisfied by $y(z)$, and hence the  ODE itself, is 
only determined in the limit. 

Whenever the sequence of approximants $E^{(n)}(x)$ converges to $E(x)$  satisfying \eqref{painleve2} and \eqref{BCs2},
as it appears to do in a surprisingly wide variety of cases [Bracken and Bass 18],
it follows that the sequence of approximants $y_E^{(n)}(z)$ converges to $y(z)$ satistfying \eqref{painleve1}
for some corresponding set of values for $a$, $b$ and $C$.

\section{Illustrative numerical examples} 

\noindent
[{\em Remark}:
Numerical approximations 
have been used for all  functions involved in the figures 
appearing here and above, and  in the evaluation of the error measure
\eqref{Delta_def}. These approximations were 
obtained using commercial packages [MATLAB 16] to solve the BVP
\eqref{painleve2}, \eqref{BCs2}, and to evaluate the integrals and Airy functions in the formulas \eqref{airy4}, \eqref{airy5} 
for $E_n(x)$.  As in [Bracken and Bass 18], the conservative view is adopted that numerical
calculations are accurate to $1$ part in $10^7$ in the determination of $E(x)$, and also of  $E^{(n)}(x)$ and
$\Delta_n$, up to $n=500$.]  

Two examples are now considered, where  BVPs 
of the form \eqref{painleve2}, \eqref{BCs2} with different parameter values lead to corresponding  BVPs of the form \eqref{painleve1}.  
The  
sequence of approximants \eqref{Esequence1}
has been considered previously for both cases 
(see the fifth and first entries in Table 1, and Figs. 5 and 3 in [Bracken and Bass 18], 
where arguments for convergence have also been presented).  

For each example, $E(x)$ is first determined, including $E(0)$ and $E(1)$, and 
then \eqref{conversion1} and \eqref{conversion2} are used to get 
the values $a$, $b$, and $C$ that complete the definition of 
the corresponding BVP \eqref{painleve1}.  
Because they are determined numerically, these values and so the BVP itself, are only known approximately.   
Also from $E(x)$,  \eqref{conversion1} and \eqref{conversion2}, a solution $y(z)$ of  \eqref{painleve1} 
is constructed.  

Next successive $E^{(n)}(x)$ are determined, including  $E^{(n)}(0)$ and  $E^{(n)}(1)$, from which  
 the values $a_n$, $b_n$, and $C_n$ are determined
using \eqref{conversion3} and \eqref{conversion4}.
Also from $E^{(n)}(x)$, successive approximants $y_E^{(n)}(z)$ to $y(z)$ are determined using \eqref{conversion3}, 
with each 
approximant defined on its 
corresponding interval $[a_n\,,b_n]$.    

The first example concerns the BVP \eqref{painleve2}, \eqref{BCs2} with
parameter values
\bea
\sigma=1/3\,,\quad \tau=-0.2\,,\quad \nu=3.5\,,\quad \mu=2.0\,.
\label{numu2}
\eea
 This BVP is known [Bracken and Bass 18] to have a unique solution $E(x)$ of Type A, as
shown in Fig. 3,  where plots of $E(x)$ and 
$E^{(n)}(x)$ are shown,  for $n=1,\,2,\,3$ on the left, and for $n=7,\,8,\,9$ on the right, suggesting convergence.  
(Note the different scales on the vertical axes for the two sets of plots.)
This suggestion is reinforced  by Fig. 4 on the  left and in the center, showing
the decrease of $\log_{10}(\Delta_n)$  to less than  $-7$ by $n=43$ and beyond,  out to $n=500$, and on the right,
the corresponding
agreement of $E^{(44)}(x)$ and $E(x)$ to better than $1$ part in $10^7$.

\begin{figure}[h]
\centering
\includegraphics[trim=0.1in 3.0in 0.1in 3.5in, clip, scale=0.6]{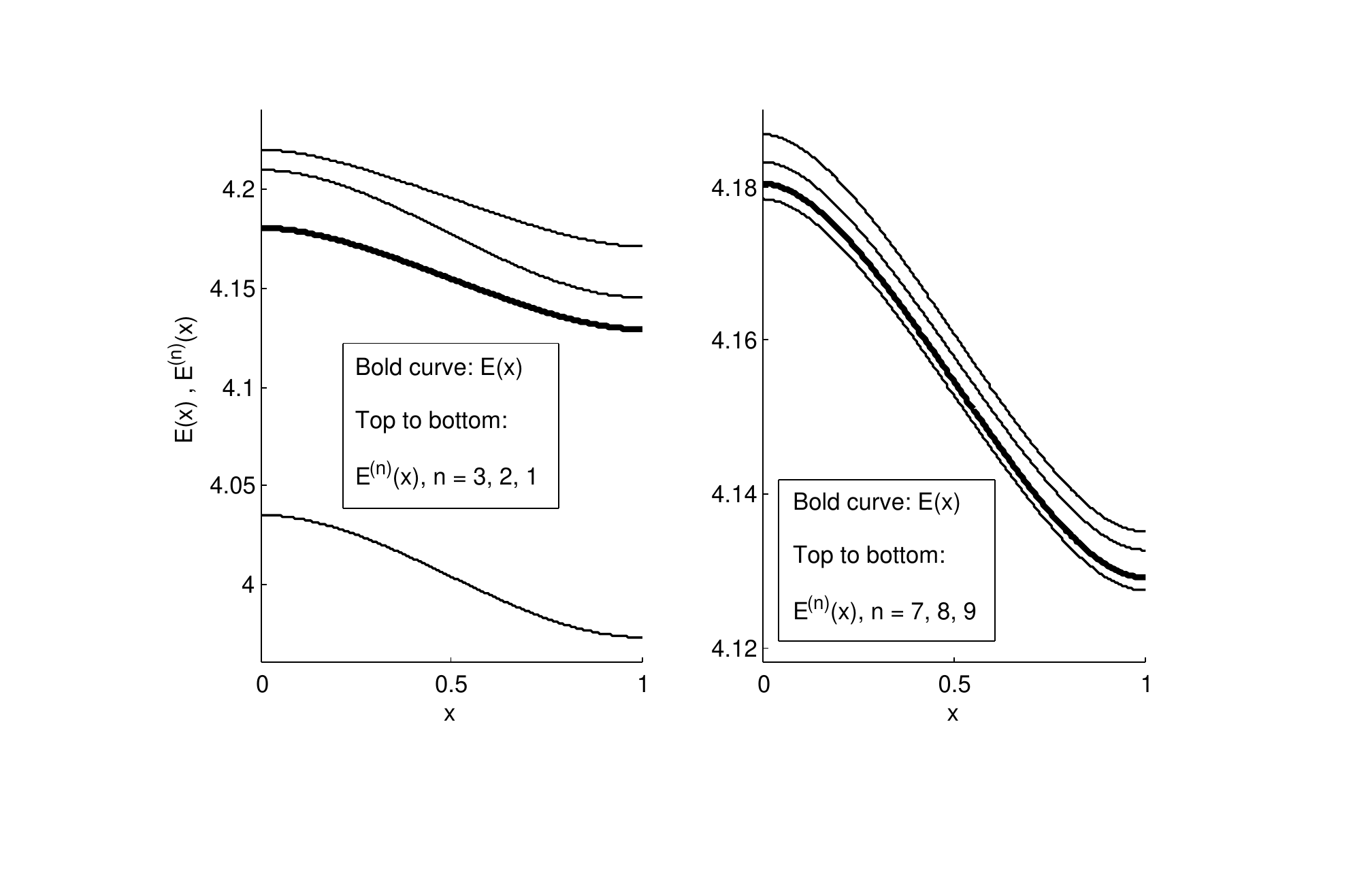}
\caption{The case $\sigma=1/3$, $\tau=-0.2$, 
$\nu=3.5$, $\mu=2.0$. Plots of $E(x)$ and $E^{(n)}(x)$ for selected values of $n$, suggesting convergence.}
\end{figure}

 \begin{figure}[h]
\centering
\includegraphics[trim=0.1in 3.0in 0.1in 3.5in, clip, scale=0.6]{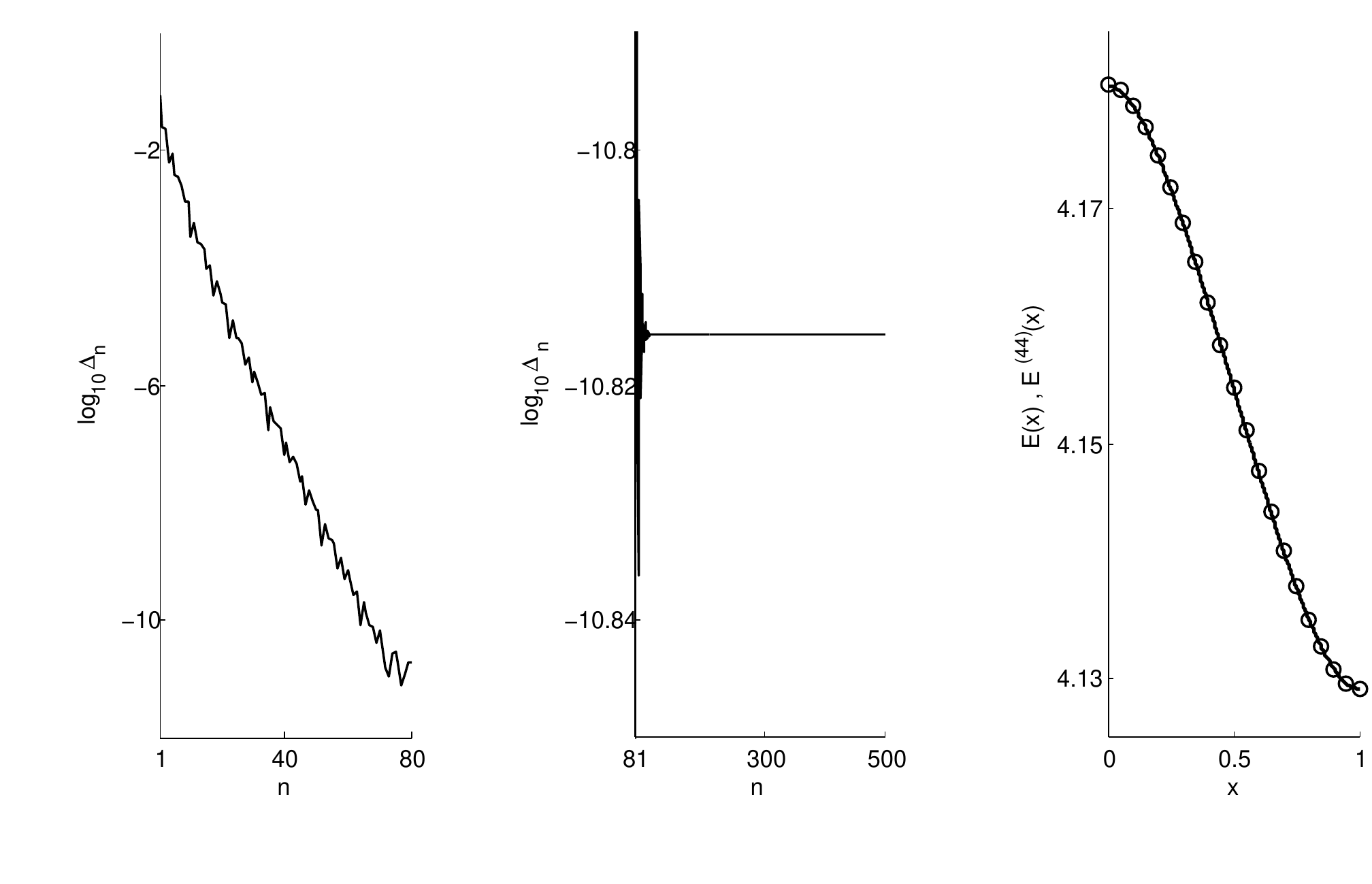}
\caption{ The case   $\sigma=1/3$, $\tau =-0.2$, $\nu=3.5$, $\mu=2.0$. On the left and in the centre, plots of $\log_{10}\Delta_n$ versus $n$, out to $n=500$.  On the right, plot of  $E(x)$ (solid line)
and $E^{(44)}(x)$ (circles) showing strong agreement. }
\end{figure}

For this example, it is found that $E(0)= 4.180\dots$ and $E(1)=4.129\dots $ (see Fig. 4), leading to
$a$, $b$ and $C$ as in \eqref{abc1}. 
That these values are only 
known approximately is now seen to be 
a result of the numerical  manipulations.  
The  solution $y(z)$ of \eqref{painleve1}, \eqref{abc1} constructed from $E(x)$ using
\eqref{conversion1} and \eqref{conversion2} is shown in Fig. 1,
together with $y_E^{(n)}(z)$ for $n=1,\,2,\,3,\,4 $ and for $n=8,\,9,\,10,\,11$, as constructed from $E^{(n)}(x)$. 
 Convergence of the $y_E^{(n)}(z)$ to $y(z)$ will follow
from convergence of the $E^{(n)}(x)$ to $E(x)$,
assuming that 
convergence  holds in the latter case as strongly suggested by the numerical computations.   

The parameter 
values $a_n$, $b_n$ and $C_n$, constructed from  $E^{(n)}(0)$ and  $E^{(n)}(1)$ using \eqref{conversion3},
\eqref{conversion4} and \eqref{conversion5}, are shown in
Fig.2. 

Note the very different character of the approach of the  extraordinary sequence of approximants
$y_E^{(n)}(z)$ to $y(z)$,
as shown in Fig. 1 
when compared with the approach of the (ordinary) sequence of approximants 
$E^{(n)}(x)$ to $E(x)$,
as shown in Fig. 3.

The second example 
concerns the BVP  \eqref{painleve2}, \eqref{BCs2}  with
parameter values 
\bea
\sigma=1/3\,,\quad \tau=-0.2\,,\quad \nu=0.1\,,\quad \mu=-0.5\,,
\label{numu3}
\eea
for which there is known to exist a unique solution of Type B.   The corresponding BVP \eqref{painleve1} in this case 
is found to have
\bea
a= 1.645\dots\,,\quad b= 3.554\dots\,,\quad C= 0.714\dots\,.
\label{abc2}
\eea

 Fig. 5 shows a plot of $y(z)$, together with $y_E^{(n)}(z)$ for $n=1\,,2$.   
In this case, apparent convergence is much more rapid than in the previous example.  At   the level of resolution
 in the figure the curve for $n=2$ is
difficult to distinguish, and
curves for larger values of $n$ cannot be distinguished at all, from the curve for $y(z)$.
 
\begin{figure}[h]
\centering
\includegraphics[trim=0.1in 3.0in 0.1in 3.5in, clip, scale=0.6]{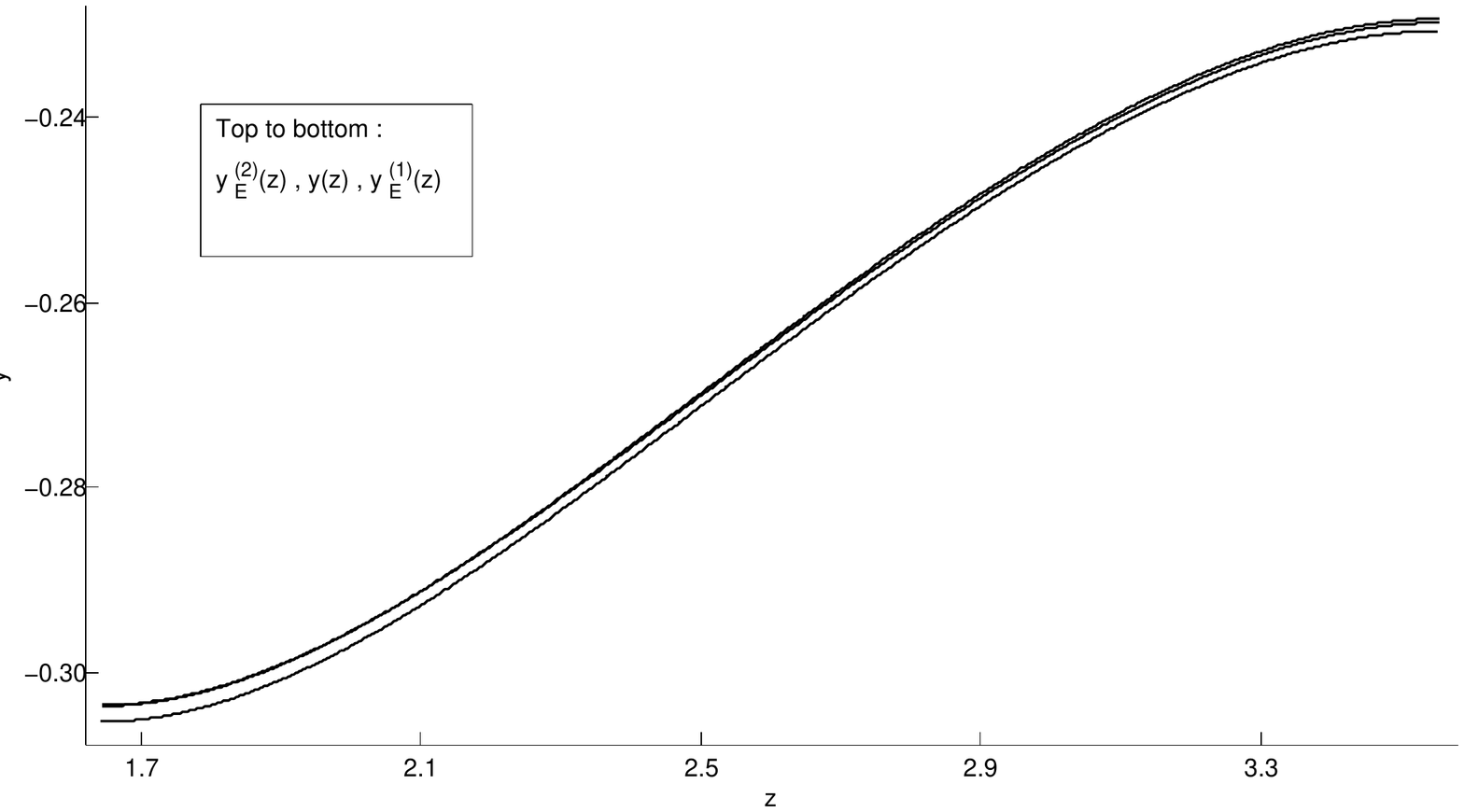}
\caption{ The case   $\sigma =1/3$, $\tau=-0.2$, 
$\nu=0.1$, $\mu=-0.5$. Plots of $y(z)$,  $y_E^{(1)}(z)$ and $y_E^{(2)}(z)$, suggesting convergence. }
\end{figure}

The different intervals of support for $y_E^{(1)}(z)$ , $y_E^{(2)}(z)$  and $y(z)$ 
 are barely discernible in Fig. 5, but (apparent) convergence of $a_n$, $b_n$ and $C_n$ values  to 
 $a$, $b$ and $C$ as in \eqref{abc2} is shown in Fig. 6.
   \begin{figure}[h]
\centering
\includegraphics[trim=0.1in 3.0in 0.1in 3.5in, clip, scale=0.5]{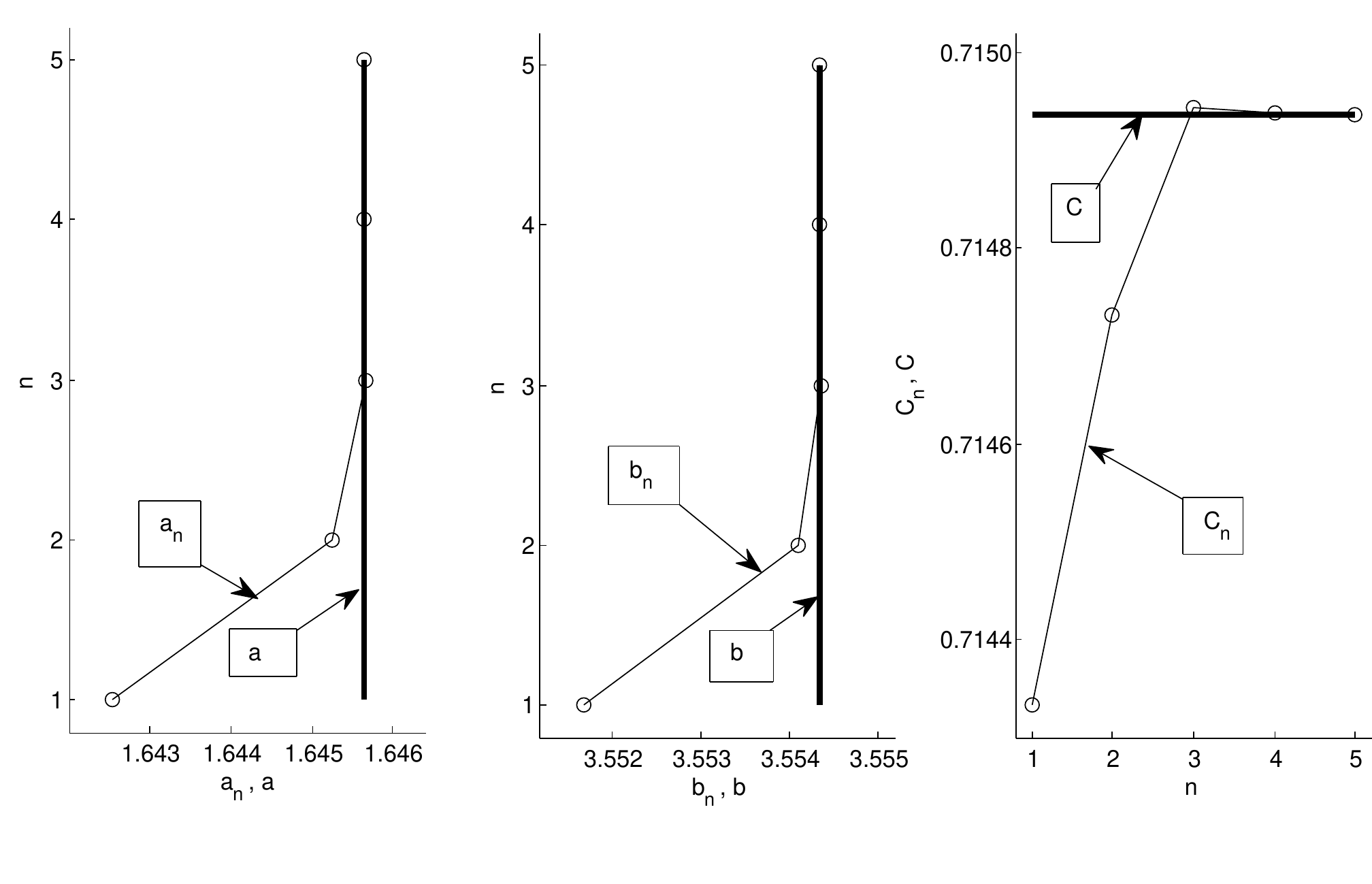}
\caption{ The case   $\sigma =1/3$, $\tau=-0.2$, 
$\nu=0.1$, $\mu=-0.5$.  Plots of $a_n$, $b_n$ and $C_n$ showing convergence to $a$, $b$ and $C$ as in \eqref{abc2}. }
\end{figure} 
The (apparent) rapid convergence of the extraordinary sequence of approximants to $y(z)$ in this case
follows from the (apparent) rapid convergence
of the corresponding sequence of approximants $E^{(n)}(x)$ to $E(x)$, as seen in Fig. 7, which shows
$\log_{10}(\Delta_n)$  decreasing to less than  $-7$ by $n=7$ and beyond,  out to $n=500$, and accordingly
the strong agreement of
$E(x)$ and $E^{(8)}(x)$ to better than $1$ part in $10^{7}$.

 \begin{figure}[h]
\centering
\includegraphics[trim=0.1in 3.0in 0.1in 3.5in, clip, scale=0.6]{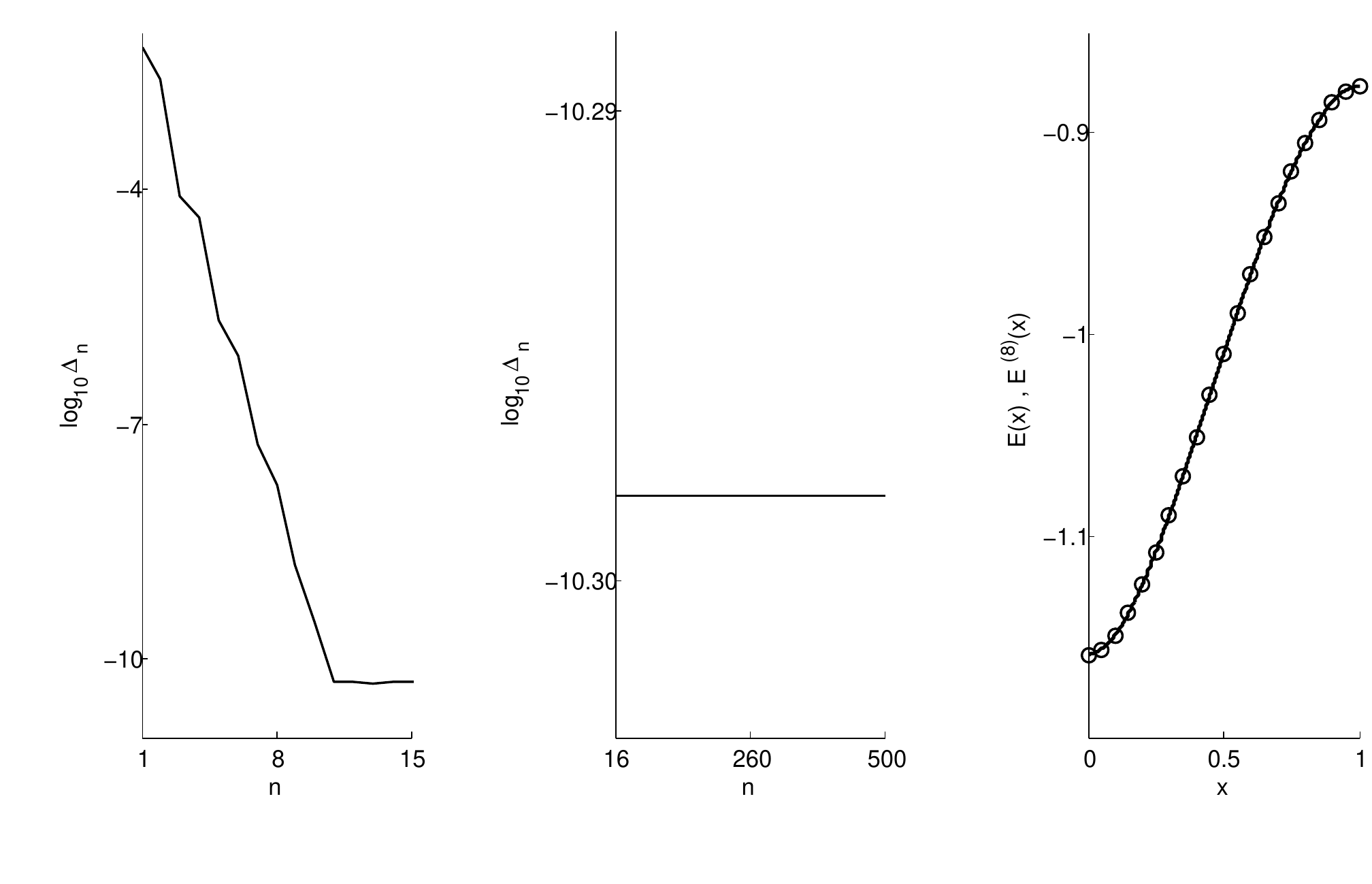}
\caption{  The case   $\sigma =1/3$, $\tau=-0.2$, 
$\nu=0.1$, $\mu=-0.5$.  
On the left and in the centre, plots of $\log_{10}\Delta_n$ versus $n$, out to $n=500$.  On the right, plot of  $E(x)$ (solid line)
and $E^{(8)}(x)$ (circles) showing strong agreement.}
\end{figure}

A perplexing feature of the construction of extraordinary sequences of approximants is 
the inability to determine {\em a priori} the appropriate parameter values in the BVP  
\eqref{painleve2}, \eqref{BCs2} that will lead to
to a given choice of $a$, $b$ and $C$ in \eqref{painleve1}.  It is the converse procedure that is more direct: Choose
parameters in \eqref{painleve2}, \eqref{BCs2}, solve that BVP numerically, and use the solution 
to determine values
for the parameters in a BVP \eqref{painleve1}.
Then determine the $E^{(n)}(x)$ and hence approximate values for $C_n$ and for the intervals
$[a_n\,,b_n]$ on which  successive approximants $y_E^{(n)}(z)$ to the solution of \eqref{painleve1} are defined.

Given that, and given that it is not known {\em a priori} for which values of $a$, $b$ and $C$ \eqref{painleve1} posseses
a singularity-free solution,  it becomes important to determine 
those values of these constants that correspond to values of the parameters appearing
in \eqref{painleve2},  in the ranges \eqref{parameters1}, 
because each such choice of those
parameters 
 is known to determine a singularity-free solution of \eqref{painleve2}, \eqref{BCs2}, from which
a corresponding
singularity-free solution of \eqref{painleve1} follows using \eqref{conversion1} and \eqref{conversion2}.  
 Unfortunately, here  it must be recognized that $a$, $b$ and $C$ must properly be regarded as  functions
of all four parameters appearing in \eqref{painleve2}.  Ideally these three functions should be evaluated over
the full
four-dimensional space of points $(\sigma\,,\tau\,,\nu\,,\mu)$ defined by \eqref{parameters1}, 
perhaps restricted as in \eqref{parameters2}, 
to determine for which BVPs \eqref{painleve1},
(apparently) convergent extraordinary approximating sequences can be found.  Unfortunately, it is difficult to explore
such  a four-dimensional space by numerical means: there is  ``a tyranny of dimensionless parameters" 
[Montroll and Shuler 79].  

The variation of $a$, $b$ and $C$  can at least be found when three
of the parameters $(\sigma\,,\tau\,,\nu\,,\mu)$ are held fixed, while one, say $\mu$, is allowed to vary.
Fig. 8 shows plots of $a$, $b$ and $C$ values determined by varying $\mu$ values for 
$\sigma=1/3$, $\tau=-0.2$ and $\nu=3.5$, including those for the first example considered above, corresponding to 
$\mu = 2.0$, shown as circled points.  Similarly, Fig. 9 shows corresponding plots when 
$\sigma=1/3$, $\tau=-0.2$ and $\nu=0.1$, including those for the second example considered above, corresponding to 
$\mu = -0.5$, again shown as circled points. 

  It can also be noted again  that when $\mu=0$, the solution of \eqref{painleve2}
and \eqref{BCs2} is $E(x)=0$.  Then \eqref{conversion1} gives $C=0$ and $y(z)=0$, 
which is the solution of \eqref{painleve1} whatever the values of $a$ and $b$ when $C=0$.     
As $\mu$ approaches $0$, and so also $E(x)$, it follows from \eqref{conversion1} that 
\bea
a\rightarrow (1-\sigma)/(4\nu\sigma^2)^{1/3}\,,\quad b\rightarrow a+(2\sigma/\nu)^{1/3}\,.
\label{ab_vals}
\eea 
With $\sigma=1/3$ this gives  
$a=0.575\dots$, $b=1.150\dots$ when $\nu=3.5$, and $a=1.882\dots$, $b=3.764\dots$ when $\nu=0.1$, providing checks on Figs. 8 and 9 at the points where $\mu=0$.  
 \begin{figure}[h]
\centering 
\includegraphics[trim=0.1in 3.0in 0.1in 3.5in, clip, scale=0.6]{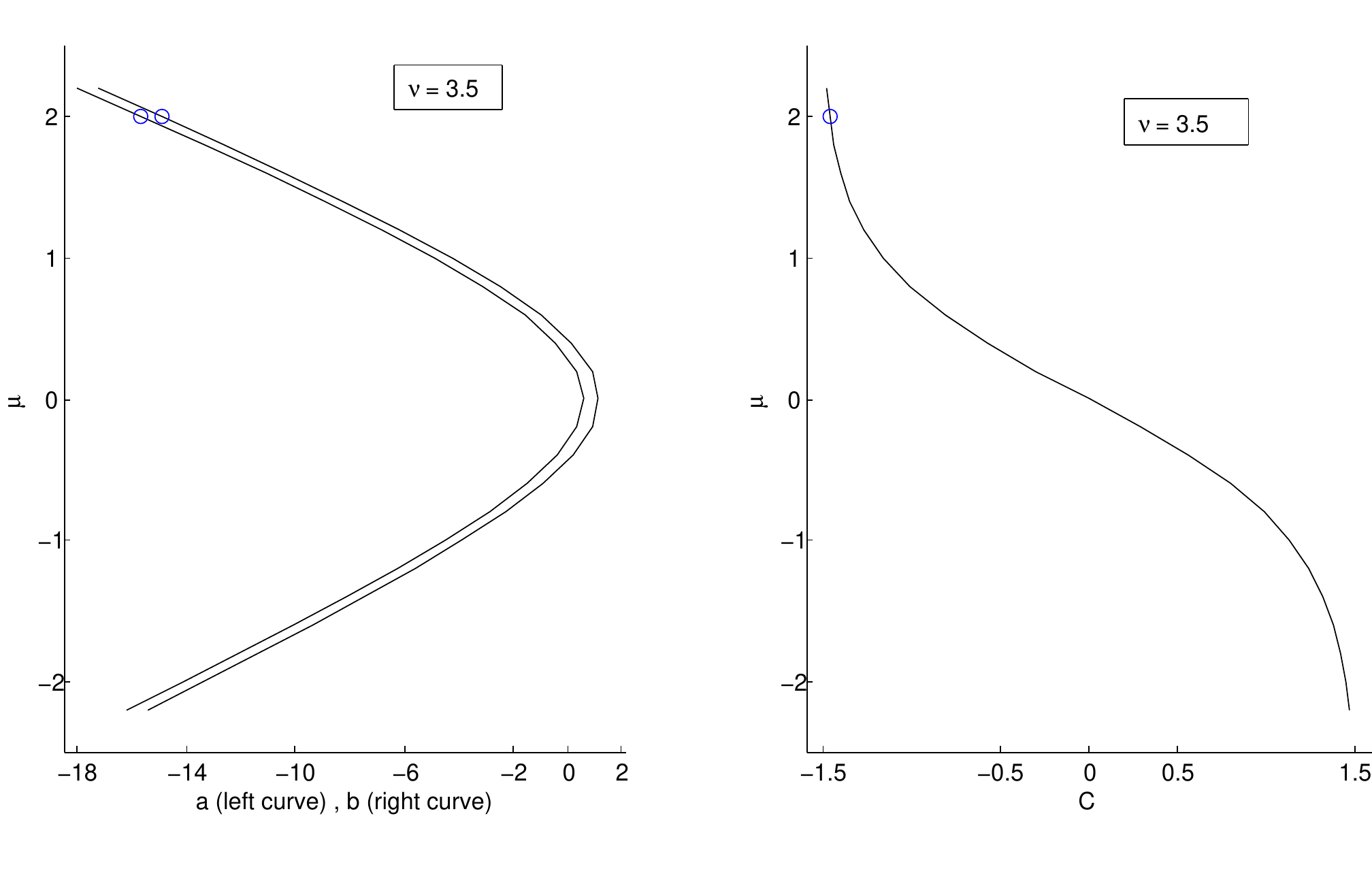}
\caption{ Ranges of $a$, $b$ and $C$ values occuring for   $\sigma=1/3$, $\tau=-0.2$, $\nu=3.5$ and variable $\mu$.
Values at $\mu=2.0$ are circled.}
\end{figure}

 \begin{figure}[h]
\centering
\includegraphics[trim=0.1in 3.0in 0.1in 3.5in, clip, scale=0.6]{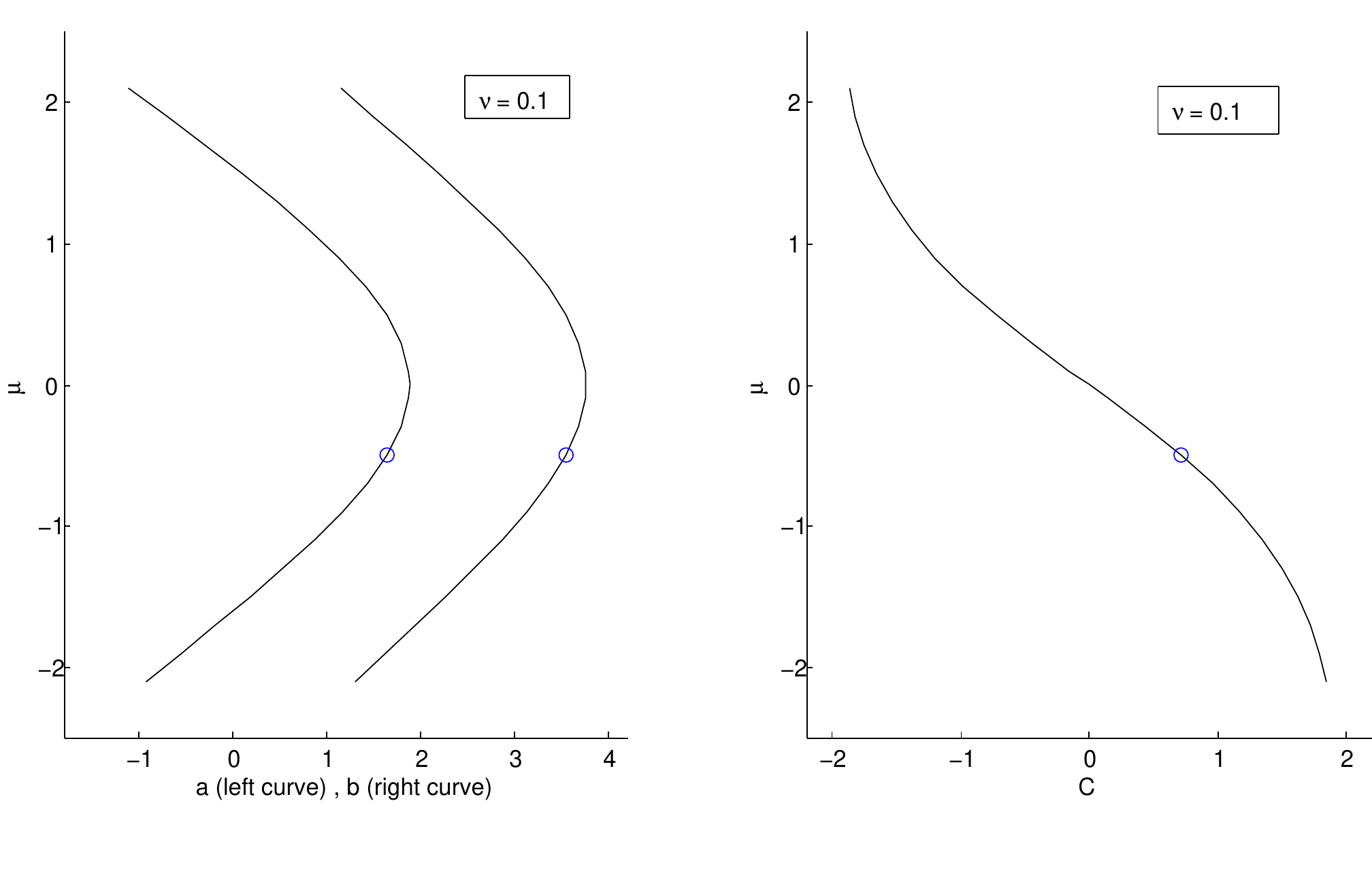}
\caption{ Ranges of $a$, $b$ and $C$ values occuring for   $\sigma=1/3$, $\tau=-0.2$, 
$\nu=0.1$ and variable $\mu$. Values at $\mu=-0.5$ are circled. }
\end{figure}

\section{Concluding remarks}
Complicated approximating sequences to solutions of BVPs may seem of little value when fast and accurate
numerical solutions are readily attainable.  The real value of the extraordinary sequences described above 
lies not in their utility as approximating schemes, but in their very existence as explicit expansions of Painlev\'e transcendents,
whose properties are still very much being explored,
in terms of familiar Airy functions.  Connections of PII with Airy functions have been obtained before 
[Lukashevich 71; Ablowitz and Segur 77; Clarkson 16],
in particular in sequences of special solutions, as mentioned earlier, 
but the expansions obtained here are very different in character from earlier results.

Perhaps the most remarkable aspect of the construction of extraordinary sequences is the apparent need to involve the
unusual supplementary ODE \eqref{painleve2} and its solutions.  It seems to be impossible to  
construct these sequences by working entirely in the framework of the BVP \eqref{painleve1}, so avoiding this 
indirect approach. 

The reader may well wonder if a more obvious  approach would  provide 
a much simpler and  more direct
way to obtain sequences of approximants to solutions of  \eqref{painleve1},  
by perturbing away from the trivial solution which applies when $C=0$.      
This approach was considered,  first replacing $C$ by $\epsilon C$ in \eqref{painleve1}
with 
the introduction of a book-keeping parameter $\epsilon$, and then expanding $y(z)$ as
\bea
y(z)=0+\epsilon y_1(z)+\epsilon^2 y_2(z)+\dots\,.
\label{pert1}
\eea
Substituting \eqref{pert1}  in the ODE and equating powers of $\epsilon$  leads to a linear BVP for each successive
$y_n(z)$, one that is explicitly solvable, again in terms of Airy functions.   A sequence of approximants can then be defined by 
\bea
y^{(n)}(z)=\sum_{k=1}^n\,y_n(z)\,,\quad n=1\,,2\,,\dots
\label{pert2}
\eea

It was found that this sequence 
apparently converges rapidly to a solution in the second case considered above, defined by \eqref{painleve1} and 
\eqref{abc2}.  However, in the first case, defined by  \eqref{painleve1} and 
\eqref{abc1}, the sequence apparently diverges.
 
It is not hard to pinpoint a key difference between the two cases.  From Fig. 5 it can be seen that in the second case
$0.23<|y(z)|<0.33$ for all $a<z<b$.  It follows that in this case
\bea
|2y(z)^3|\ll |zy(z)|\,,\quad |2y(z)^3|\ll |C|\,,\quad a<z<b\,,
\label{y_size1}
\eea
so the nonlinear term in the ODE is everywhere small compared with the other terms on the RHS, and it is no surprise that
the perturbation sequence (apparently) converges.  
In contrast, Fig. 1 shows that in the first case $|y(z)|>2.79$ for all $a<z<b$, 
so the nonlinear term is not small compared with the other terms on the RHS and the sequence (apparently) diverges. 

It is quite unclear why the much more complicated approach described in Sec. 2 produces an apparently 
convergent sequence, especially in cases like
that in the first example above where ordinary perturbation fails.  Indeed, it remains a mystery why
the sequence of functions $E^{(n)}(x)$ in \eqref{Esequence1} apparently converges to the solution $E(x)$ of the BVP 
\eqref{painleve2}, \eqref{BCs2} in such a wide variety of cases [Bracken and Bass 18], leading to apparently convergent 
extraordinary sequences of approximants to solutions of BVPs \eqref{painleve1}  in a correspondingly wide range of cases.

 Evidently the results presented above, being largely based on numerical experiments,  
 pose important unanswered questions warranting further analysis. But such analysis is 
 beyond the scope of the present work.   
\vskip5mm\noindent   
{\bf {\Large Acknowledgement:}} The author thanks  Ludvik Bass for many 
stimulating conversations, and for introducing him to the abundance of interesting mathematical
problems associated with BVPs of the form \eqref{painleve2}, \eqref{BCs2}.

\vskip5mm\noindent  
{\bf {\Large References}}
\vskip5mm\noindent   
[Ablowitz and Segur 77]
M.J. Ablowitz and H. Segur.  ``Exact Linearization of a Painlev\'e Transcendent."  {\em Phys. Rev. Letts.}  38 (1977), 1103--1106. 
\vskip5mm\noindent   
[Abramowitz and Stegun 64]
M. Abramowitz and I.A. Stegun (Eds.). {\em Handbook of Mathematical Functions}. 
New York: Dover, 1964, Ch. 10.  
\vskip5mm\noindent   
[Amster et al. 11]
P. Amster,  M.K. Kwong and C. Rogers. 
``On a Neumann Boundary Value Problem for the Painleve II Equation in Two-Ion Electro-Diffusion." 
{\em Nonlin. Anal. Th. Meth. App.}    74 (2011),  2897-2907. 
\vskip5mm\noindent   
[Bass 64]
L. Bass.  ``Electrical structures of interfaces in steady electrolysis."  {\em Trans. Faraday Soc.}  60 (1964),  1656--1663. 
\vskip5mm\noindent   
[Bass et al. 10]
L. Bass, J.J.C. Nimmo,  C. Rogers and W.K. Schief.  ``Electrical Structures of Interfaces: A Painlev\'e II Model." 
{\em Proc. Roy. Soc. (London) A} 466 (2010),   2117--2136. 
\vskip5mm\noindent   
[Bass and Bracken 14]
L. Bass and A.J. Bracken. ``Emergent Behaviour in Electrodiffusion: Planck's Other Quanta."  
{\em Rep. Math. Phys.}  73 (2014), 65--75.
\vskip5mm\noindent   
[Bassom et al. 98]
A.P. Bassom,  P.A. Clarkson,  C.K. Law and J.B. McLeod. 
``Application of Uniform Asymptotics to the Second Painlev\'e Transcendent." 
 {\em Arch. Rat. Mech. Anal.}  143 (1998), 241--271.
\vskip5mm\noindent   
[Bracken et al. 12]
A.J. Bracken, L. Bass and C. Rogers.  ``B\"acklund Flux-Quantization in a Model of Electrodiffusion Based on Painlev\'e II." 
{\em J. Phys. A: Math. Theor.} 45 (2012), 105204.
\vskip5mm\noindent   
 [Bracken and Bass 16]
  A.J. Bracken and L. Bass. 
``Differential Equations of Electrodiffusion: Constant Field Solutions, Uniqueness, 
and  New Formulas of Goldman-Hodgkin-Katz Type."  
{\em SIAM J. App. Math.} 76 (2016), 2286--2305.
\vskip5mm\noindent   
[Bracken and Bass 18]
A.J. Bracken and L. Bass. ``Series Solution of Painlev\'e II in Electrodiffusion: Conjectured Convergence."
{\em J. Phys. A: Math. Theor.} 51 (2018), 035202.
\vskip5mm\noindent   
 [Clarkson 06]
 P.A. Clarkson.  {\em Painlevé Equations: Nonlinear Special Functions.}  Orthogonal Polynomials and Special Functions: Computation and Application  
(F. Márcellan  and W. van Assche, eds), Lecture Notes  in Mathematics, vol. 1883, Berlin: Springer-Verlag, 2006, pp. 331--411 . 
\vskip5mm\noindent   
[Clarkson 16]
P.A. Clarkson.  ``On Airy Solutions of the Second Painlev\'e Equation."  {\em Stud. App. Math.} 137 (2016), 93--109.
\vskip5mm\noindent 
[Ince 56]
E.L. Ince. {\em Ordinary Differential Equations}. New York: Dover, 1956. 
\vskip5mm\noindent  
[Joshi and Kruskal 94]
N. Joshi and M.D. Kruskal. 
``A Direct Proof That Solutions of the 6 Painlev\'e  Equations Have No Movable Singularities Except Poles." 
{\em Stud. Appl. Math.}    93 (1994), 187--207.
\vskip5mm\noindent   
[Kajiwara et al. 17]
K. Kajiwara, M. Noumi and Y. Yamada. ``Geometric Aspects of Painlev\'e Equations."
{\em J. Phys. A: Math. Theor.} 50 (2017), 073001.
\vskip5mm\noindent   
[Lukashevich 71]
N.A. Lukashevich. ``The Second Painlev\'e Equation."  {\em Diff. Equ.}  7 (1971), 853--854.  
\vskip5mm\noindent   
[MacGillivray 68]
A.D. MacGillivray. ``Nernst-Planck Equations and Electroneutrality and Donnan Equilibrium Assumptions."
{\em J. Chem. Phys.} 48 (1968) 2903--2907.
\vskip5mm\noindent   
[MATLAB 16]
{\em MATLAB} Mass. USA: MathWorks, 2016. 
\vskip5mm\noindent   
[Montroll and Shuler 79]
E.W. Montroll and K.E. Shuler. 
``Dynamics of Technological Evolution -- Random-Walk Model for the Research Enterprise." 
 {\em Proc. Natl. Acad. Sc. USA}  76 (1979), 6030--6034. 
\vskip5mm\noindent   
[Painlev\'e 02]
P. Painlev\'e.  
``Sur les \'Equations Diff\'erentielles du Second Ordre et d'Ordre Sup\'erieur dont l'Int\'egrale G\'en\'erale est Uniforme." 
{\em Acta Math.} 25 (1902), 1--85.
\vskip5mm\noindent   
[Park and Jerome 97]
J.-H. Park and J.W. Jerome. 
``Qualitative Properties of Steady-State Poisson-Nernst-Planck  Systems: Mathematical Study." 
{\em SIAM J. Appl. Math.} 57 (1997), 609--630.
\vskip5mm\noindent   
[Rogers et al. 99]
C. Rogers,  A.P. Bassom and W.K. Schief.  ``On a Painleve II Model in Steady Electrolysis: Application of a B\"acklund Transformation." 
{\em J. Math. Anal. App.}    240 (1999),    367--381.
\vskip5mm\noindent   
[Sakai 01] 
H. Sakai. ``Rational Surfaces Associated with Affine Root Systems and Geometry of the Painlev\'e Equations."
{\em Commun. Math. Phys.} 220 (2001), 165--229.
\vskip5mm\noindent   
[Thompson 94]
 H.B. Thompson. ``Existence for 2-Point Boundary-Value-Problems in 2-Ion Electrodiffusion."  
 {\em J. Math. Anal. App.} 184 (1994), 82--94.
 \vskip5mm\noindent   
 [Vorob'ev 65]
 A.P. Vorob'ev. ``On Rational Solutions of the Second Painlev\'e Equation."
 {\em Diff. Equ.}  1 (1965), 79--81. 
 \vskip5mm\noindent   
 [Yablonskii 59]
 A.I. Yablonskii.   ``On Rational Solutions of the Second Painlev\'e Equation."
 {\em Vesti. Akad. Nauk BSSR. Ser. Fiz. Tkh. Nauk} 3 (1959), 30--35. 
 \vskip5mm\noindent   
 [Zaltzman and Rubinstein 07]
 B. Zaltzman and I. Rubinstein. ``Electro-Osmotic Slip and Electroconvective Instability."
 {\em J. Fluid Mech.} 579 (2007), 173--226

\end{document}